\documentclass{aa}
\usepackage{graphicx}
\begin{document}
    \title{Intrinsic spectral blueshifts in rapidly rotating stars?}
    \author{S{\o}ren Madsen, Dainis Dravins, Hans-G\"{u}nter Ludwig and Lennart Lindegren}
    \offprints{S. Madsen, {\tt soren@astro.lu.se}}
    \institute{
    Lund Observatory, Box~43,
    SE--22100 Lund, Sweden\\
    \email{soren, dainis, hgl, lennart@astro.lu.se}}

\date{Received -- -- -- / Accepted -- -- --}
\titlerunning{Blueshifts in rapidly rotating stars?}
\authorrunning{Madsen et al.}

\abstract{
Spectroscopic radial velocities for several nearby open clusters suggest that spectra of (especially earlier-type) rapidly rotating stars are systematically blueshifted by 3~km~s$^{-1}$ or more, relative to the spectra of slowly rotating ones.  Comparisons with astrometrically determined radial motions in the Hyades suggests this to be an absolute blueshift, relative to wavelengths naively expected from stellar radial motion and gravitational redshift.  Analogous trends are seen also in most other clusters studied (Pleiades, Coma Berenices, Praesepe, $\alpha$ Persei, IC 2391, NGC 6475, IC 4665, NGC 1976 and NGC 2516).  Possible mechanisms are discussed, including photospheric convection, stellar pulsation, meridional circulation, and shock-wave propagation, as well as effects caused by template mismatch in determining wavelength displacements.  For early-type stars, a plausible mechanism is shock-wave propagation upward through the photospheric line-forming regions.  Such wavelength shifts thus permit studies of certain types of stellar atmospheric dynamics and -- irrespective of their cause -- may influence deduced open-cluster membership (when selected from common velocity) and deduced cluster dynamics (some types of stars might show fortuitous velocity patterns).

\keywords{
Methods: data analysis --
Techniques: astrometry -- spectroscopy -- radial velocities --
Stars: rotation --
Open clusters and associations: general}}

\maketitle


\section{Introduction}

Wavelength displacements in stellar spectra are affected not only by stellar radial motion causing a Doppler shift, but also by line asymmetries and shifts caused by, e.g., atmospheric pulsation, surface convection, stellar winds, and gravitational potential.  However, for solar-type stars, the net effect of convective blueshifts and gravitational redshifts, as predicted from hydrodynamic model atmospheres, does not exceed 1~km~s$^{-1}$ (e.g., Dravins \cite{dravins99}).

For a few nearby open clusters, it has been possible to determine accurate radial motions from astrometry only, not invoking spectroscopy.  Comparing such astrometric radial velocities to spectroscopic data for the Hyades, indicates not only the existence of such convective lineshifts, but also an increasing blueshift of stellar spectra, reaching 3--5~km~s$^{-1}$ for rapidly rotating stars earlier than about F5~V: Madsen et al.\ (\cite{madsen02}) and Dravins (\cite{dravins03}).  Since there do not yet exist any detailed hydrodynamic models of such earlier-type stellar atmospheres from which synthetic spectral-line shifts could be reliably computed, it is not yet possible to theoretically verify the reality of these suggested trends.

In order to examine this effect, we have analyzed spectroscopic data for as large a dataset as possible, using the open-cluster database
WEBDA\footnote{Available at http://obswww.unige.ch/webda/ }.
Although its data sets are somewhat inhomogeneous, this permits a study of relative wavelength displacements for ten clusters, as function of stellar spectral type and of rotational velocity.

Five of these clusters have astrometric radial velocities deduced from Hipparcos data (Madsen et al.\ \cite{madsen02}): Hyades, Pleiades, Coma Berenices, Praesepe and $\alpha$ Persei, but only for the Hyades (and perhaps also Coma) is the accuracy adequate to permit studies of not only relative but also {\em absolute} wavelength shifts. For stars not measured from Hipparcos, their astrometric radial velocities are estimated by assuming they follow the cluster's common space motion.

Spectroscopic radial velocities for the other five clusters (IC 2391, NGC 6475, IC 4665, NGC 1976 and NGC 2516) permit studies of {\em relative} wavelength shifts relative to their mean.  This is possible for clusters of small angular size whose stars share the same velocity vector, so that the variation across the sky of the radial motions is negligible compared to the shifts one is searching for.

All ten clusters have metallicities close to solar. Their ages range from
13~Myr (NGC 1976) to 660~Myr (Praesepe).

\section{The 10 clusters}
\label{sec:hyad}

\begin{figure}
   \resizebox{\hsize}{!}{\includegraphics*{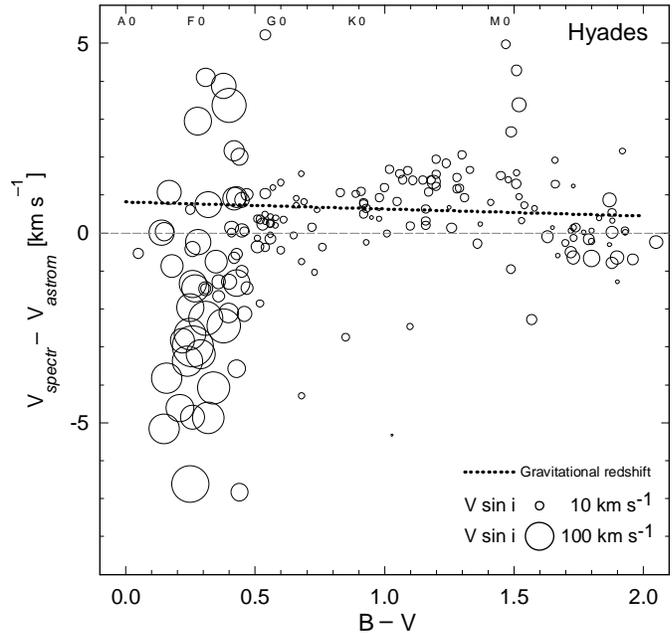}}

\caption{
Wavelength displacements of stellar spectra in the Hyades. Differences between spectroscopically measured radial velocities, and astrometrically determined radial motion are plotted as function of both $B-V$, and projected rotational velocity $V\sin i$ (proportional to symbol area). Spectra of rapidly rotating early-type stars appear systematically more blueshifted than those of slower-rotating later-type ones. Only points with nominal differential velocity errors $<$~3~km~s$^{-1}$ are plotted. In the absence of effects in stellar atmospheres or due to spectroscopic techniques, the naively expected spectral displacements should equal the marked line which estimates the gravitational redshift for main-sequence stars.  Corresponding spectral types for main-sequence stars are marked at the top.
}
\label{fig:hyad}
\end{figure}

\begin{figure*}

\centering
  \includegraphics[width=17cm]{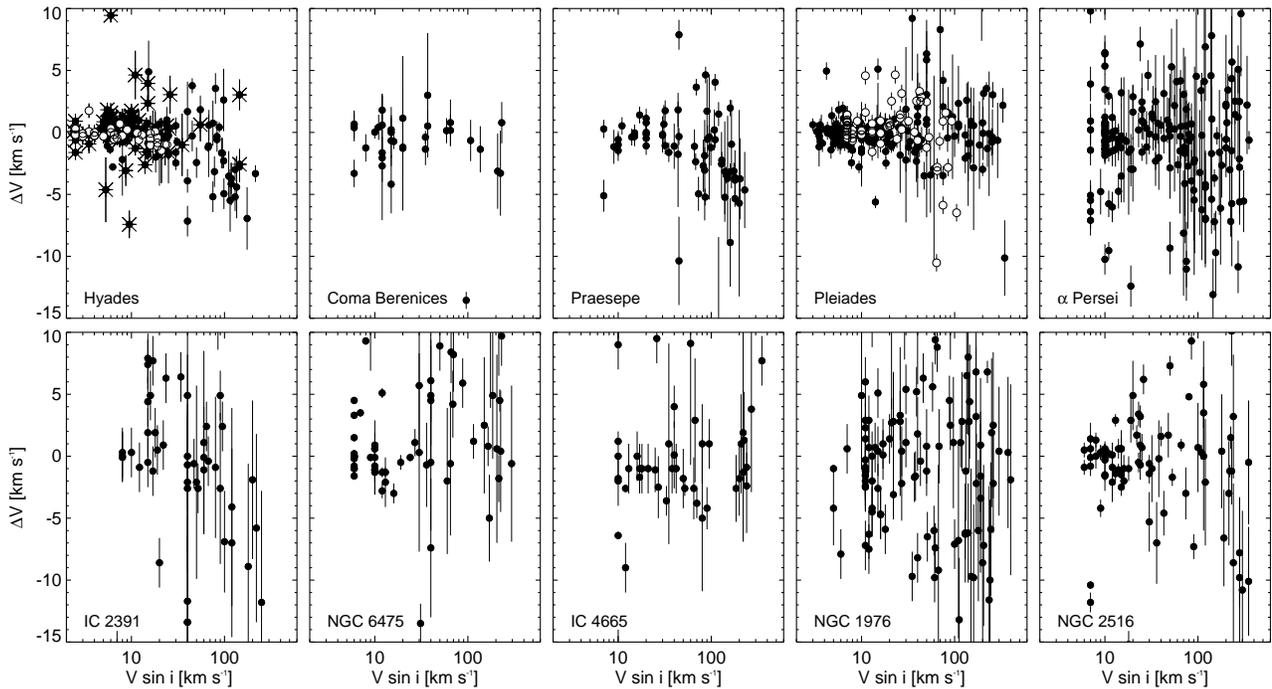}

\caption{Differences between spectroscopic and astrometric radial velocities versus stellar rotational velocity $V \sin i$ (top), and relative spectroscopic velocities for clusters without such astrometric data (bottom). For each cluster, the values are normalized to have $\Delta V = 0$ correspond to the median radial velocity of stars with $V \sin i \le$~60~km~s$^{-1}$.  The Hyades: black points are from Madsen et al.\ (\cite{madsen02}); open circles are M dwarfs (known binaries excluded) from Reid \& Mahoney (\cite{reid00}); asterisks are other assumed cluster members.  The Pleiades: Open circles are M dwarfs from Terndrup et al. (\cite{terndrup00}).  For the Hyades, stars with errors $>$~3~km~s$^{-1}$ are omitted; for other clusters those with $>$~10~km~s$^{-1}$.  Those with no errors in the literature were assigned the value 2~km~s$^{-1}$.
}
\label{fig:rot}
\end{figure*}

Fig.~\ref{fig:hyad} shows absolute wavelength shifts of stellar spectra in the Hyades, plotted as spectroscopic minus astrometric radial velocities.  A slight trend of a somewhat increased blueshift (by about 1~km~s$^{-1}$) when going from K-type to mid-F type stars ($B-V$ from about 1.1 to 0.5) is expected due to the more vigorous convection in F stars, causing increased convective blueshift.  However, this trend continues and becomes enhanced for even hotter stars, a region where not many detailed hydrodynamic models have yet been developed.  Many of these stars are also rapidly rotating, making it difficult to disentangle whether the dependence is upon spectral type or/and on rotation, or perhaps even on something else.  A similar dependence on rotation is suggested also from data for some cool Hyades M-stars: Fig.~\ref{fig:rot}.  The trend seen in the Hyades is also present in Praesepe  and possibly in Coma Berenices. In $\alpha$ Persei the same is seen except for the most rapid rotators (all of spectral type B). For the Pleiades, the trend for earlier spectral types is at best weak, while its M--stars do show a clear correlation, noted also by Terndrup et al. (\cite{terndrup00}); see their Fig.~3.  The error bars in Fig.~\ref{fig:rot} include both the stated spectroscopic errors and the part of the astrometric error coming from the internal velocity dispersion in each cluster.  For detailed discussions on the astrometric error budget, see Lindegren et al.\ (\cite{lindegren00}) and Madsen (\cite{madsen03}).

The rotational broadening of spectral line profiles makes the {\em spectroscopic} determination of wavelength shifts difficult in rapidly rotating stars, and even the errors of several km~s$^{-1}$ quoted in the literature may be optimistic.  Therefore, one should not draw conclusions from individual data points for rapid rotators in, e.g., Fig.~1, but only from their averages and from the general trends.  The statistical scatter may well account for some stars showing apparent redshifts.

The trend of increasing blueshift with higher $V \sin i$ can be seen in IC 2391 and NGC 2516, partly in IC 4665, maybe in NGC 1976, but not in NGC 6475 (Fig.~\ref{fig:rot}).  In contrast with $\alpha$ Persei and maybe the Pleiades, NGC 1976 and IC 4665, the most blueshifted stars in IC 2391 and NGC 2516 happen to be of B-type. It must be noted that all these five clusters generally lack mid- and late A stars, for which the trend is well visible in the Hyades.  There is no apparent correlation between the suggested trends and cluster age.  For instance, IC 2391 and IC 4665 are coeval (45~Myr), while the Pleiades (130~Myr) and NGC 2516 (110~Myr) have about the same age, but show different trends.

The compiled data sets of spectroscopic radial velocities and projected rotational velocities $V \sin i$ are quite inhomogeneous, especially for the early spectral types. Since the number of sources is huge, we only give some references to observations or compilations for the interesting part of Figs.~1 and 2 (i.e. where $V \sin i$ shows a trend towards blueshifted spectra).  The Hyades set is a mixture of the compilations by Perryman et al.\ (\cite{perryman98}) and G\l{}ebocki \& Stawikowski (\cite{glebocki00}), and the observations by Griffin et al.\ (\cite{griffin88}) and Reid \& Mahoney (\cite{reid00}). The Pleiades set mainly comes from Liu et al. (\cite{liu91}), Morse et al. (\cite{morse91}), Raboud \& Mermilliod (\cite{raboud98}), Queloz et al.\ (\cite{queloz98}) and Terndrup et al.\ (\cite{terndrup00}). Coma Berenices data are largely from Abt \& Willmarth (\cite{abt99}). Praesepe data are chiefly from McGee et al. (\cite{mcgee67}), Rachford (\cite{rachford98}) and Abt \& Willmarth (\cite{abt99}). Data for $\alpha$ Persei primarily come from Kraft (\cite{kraft67}), Petrie \& Heard (\cite{petrie70}), Millward \& Walker (\cite{millward85}), Prosser (\cite{prosser92}) and Morrell \& Abt (\cite{morrell92}). For IC 2391, the data are mainly from Levato \& Garcia (\cite{levato84}) and Levato et al. (\cite{levato88}); for NGC 6475 mainly from Abt \& Jewsbury (\cite{abt69}), Abt et al.\ (\cite{abt70}) and Gieseking (\cite{gieseking77}), and for IC 4665 predominantly from Abt \& Snowden (\cite{abt64}), Abt \& Chaffee (\cite{abt67}) and Morrell \& Abt (\cite{morrell91}). NGC 1976 data are for the most part from Abt et al. (\cite{abtea70}), Smith et al.\ (\cite{smith83}), Morrell \& Levato (\cite{morrelll91}) and Abt et al.\ (\cite{abt91}), while those for NGC 2516 mainly originate from Abt et al.\ (\cite{abtea69}), Abt \& Levy (\cite{abt72}), Herwig (\cite{herwig97}) and Gonz\'{a}lez \& Lapasset (\cite{gonzalez00}).

Instead of open clusters, binaries may be used to study lineshifts in stars (e.g. Pourbaix et al.\ \cite{pourbaix02}). However, studies of visual-spectroscopic binaries (Pourbaix \cite{pourbaix00}), including such with one component being a rapidly rotating A star, have not revealed trends of blueshifts correlated with stellar rotation, but the effect could be hidden in the errors of the estimated binary parameters.

\section{Spectrum shifts due to atmospheric dynamics}

Although data for different clusters are thus somewhat scattered, there still remains the suggestion of a common trend, such that more rapid stellar rotation (and early spectral type) correlates with increased spectral blueshift.  In this Section, we now look at some plausible mechanisms that might explain such shifts.

\subsection{Convection}

Convective blueshifts of spectral lines occur in solar-type stars due to the correlation between upward motions of photospheric gas, and their local brightness.  Rising and blueshifted elements of hot (bright) gas contribute a larger number of photons than the same gas when it has cooled off (is darker) and is sinking (thus redshifted).  The effect for the Sun is typically 300--400~m~s$^{-1}$, increasing to about 1~km~s$^{-1}$ in hotter F--type stars (Allende Prieto et al. \cite{allende02b}).

\begin{figure}
  \includegraphics[width=8.9cm]{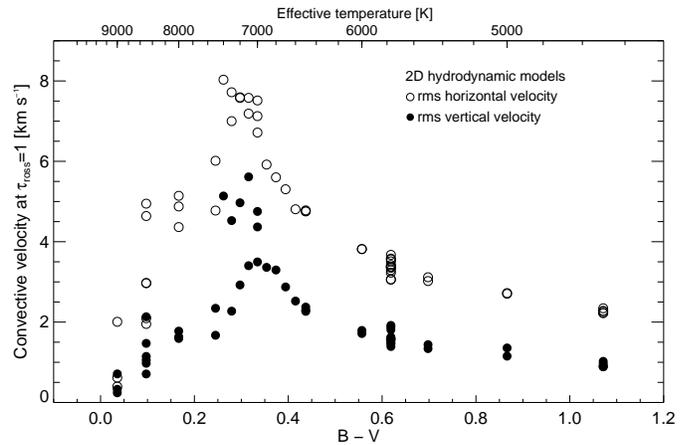}
\caption{Convective velocities from a grid of 55 main-sequence hydrodynamical model atmospheres (Ludwig et al. \cite{ludwig99}). Shown is the root-mean-square horizontal (open) and vertical (solid circles) convective velocity at Rosseland optical depth unity, as function of $B-V$. The corresponding effective-temperature scale is at the top.}
\label{fig:vconv}
\end{figure}

The effects suggested by Figs.~\ref{fig:hyad} and \ref{fig:rot}, however, are of greater magnitude, and are unlikely to be explained by convective blueshifts.  This assessment is based on a theoretical grid of 55 two-dimensional hydrodynamic model atmospheres (Ludwig et al. \cite{ludwig99}) for A- to K-type main-sequence stars with solar metallicity.  Fig.~\ref{fig:vconv} shows the run of the horizontal and vertical component of the convective velocity at Rosseland optical depth unity, as function of $B-V$.  For the conversion from effective temperature to color, a semi-empirical relation in Alonso et al.\ (\cite{alonso96}) was used.  Some scatter is apparent: the large velocities in six models around $B-V=0.3$ are due to radial oscillations which are not of convective origin, and maximum convective velocities of about 2.2 times solar are found at $B-V=0.33$ (7000\,K).  For a given stellar surface contrast, the velocity scaling should be indicative of also the scaling for line shifts.  For mid F-type dwarfs some enhancement occurs due to an increased granular temperature contrast but even then the convective blueshift would not exceed about 1~km~s$^{-1}$, at least not in the optical wavelength region.  Velocities and temperature contrasts quickly drop towards A-type objects, reducing the shifts.

The convective blueshift so far discussed is not correlated with stellar rotation.  However, a dependence could arise if rapid rotation would somehow modify the convective flows under the influence of Coriolis and centrifugal forces, enhancing the blueshift.

The horizontal scale of granulation seems to be proportional to the pressure scale height at optical depth unity (Freytag et al. \cite{freytag97}). The velocities of Fig.~\ref{fig:vconv} lead to convective turn-over time scales for A- to G-type dwarfs within a factor 2 of the solar value ($\approx 10$\,min).  Even at rotation rates close to break-up, turn-over time scales remain more than one order of magnitude shorter than the rotational period, rendering Coriolis forces rather ineffective for modulating the convective surface structures.  Centrifugal forces reduce the effective surface gravity, producing more ``giant-like'' conditions.  At given effective temperature, lower-gravity models tend to produce larger shifts, but observed velocity spans of line bisectors in giants hardly exceed 2~km~s$^{-1}$ (Gray \&\ Nagel \cite{gray89}, Allende Prieto et al. \cite{allende02a}).  Since large lineshifts can be expected to be accompanied by also large line asymmetries, it appears improbable that lineshifts of 3--5~km~s$^{-1}$ could be produced by effects of rotation on convection.

\subsection{Pulsation}

Gas motions where upward velocity correlates with increased brightness may cause blueshifted spectra, irrespective of the scale of the gas flows.  Quite a few among the rapidly rotating earlier-type stars showing spectral blueshifts reside in the area of the Hertzsprung-Russell diagram populated by Delta-Scuti-type and similar variables.  The {\em sign} of the effect appears correct in the sense that maximum outward motion in these pulsating stars occurs near the phase of stellar maximum brightness; thus an average over a pulsation cycle should lead to somewhat blueshifted lines.

Although the number of [known] pulsating Delta Scuti stars in open clusters is not especially great (Antonello \& Pasinetti Fracassini \cite{antonello98}; Rodr{\'\i}guez \& Breger \cite{rodriguez01}), several are rapidly rotating and might influence our statistical trends.  Detailed studies of the radial-velocity and line-shape variations have been made for a few larger-amplitude Delta Scuti stars (e.g., Dravins et al.\ \cite{dravins77}; Yang et al.\ \cite{yang87}), but larger pulsational amplitudes are confined to the slow rotators (Breger \cite{breger00}).  Since typical amplitudes in Delta Scuti stars are modest (maybe 3~km~s$^{-1}$), and typical photometric amplitudes only some per cent, these appear inadequate for causing a statistical bias of the magnitudes suggested by Figs.~\ref{fig:hyad} and \ref{fig:rot}.

\subsection{Meridional circulation}

Wavelength shifts may offer a possibility to study large-scale
stationary flows across stellar surfaces, such as {\em meridional
circulation}.  In a rotating star, meridional flows are generated with
-- in the simplest case -- rising motions at the poles and sinking at
the equator (Sweet \cite{sweet50}).  The classical theory predicts
very slow motions, occurring on the Kelvin-Helmholtz time-scale of the
star, implying slow flows even in fast rotators. However, the
classical theory becomes questionable at the flow boundaries (for a
discussion see Tassoul \cite{tassoul00}) due to divergent solutions
towards infinite flow speeds.  Hitherto, this problem has not been
clarified, not excluding the possibility that meridional flows in the
atmosphere could be faster than at first expected.  In our case, the
measured spectral-line displacements originate mainly from the upper
stellar photosphere, while the classical theory is about motions in
stellar interiors.  Issues include whether the very much lower
densities in stellar atmospheres (as compared to the interiors)
somehow might cause more rapid flows there (to perhaps conserve
momentum), whether there are counter-rotating ``cells'' near stellar
surfaces (Tassoul \& Tassoul \cite{tassoul83}; their Fig.~1), or
other dynamic issues connected with the relatively thin line-forming
layers.

Related large-scale flows at the stellar surface are thermally
 driven ``winds'': rotation leads to gravity darkening, i.e. a
 non-uniform temperature distribution, with the stellar poles hotter
 than the equatorial regions. A localized heating of an
 atmosphere was studied by Showman \&\ Guillot (\cite{showman02}) in
 the context of driving (zonal) winds on irradiated extra-solar giant
 planets. They find wind speeds up to 2~km~s$^{-1}$ for temperature
 contrasts similar to those expected on rapidly rotating stars.

The radial-velocity component of a meridional circulation pattern is
much smaller than its tangential component (e.g. Tassoul
\cite{tassoul00}). The tangential velocity is -- again in the
simplest case -- directed from the poles towards the equator of a
rotating star.  Such a flow seen equator-on would spectroscopically
produce a net blueshift, seen pole-on a net redshift.  Stars having a large $V \sin i$ are likely observed nearly equator-on, and for such stars one would expect blueshifts.  However, even if one assumes little cancellation by
the averaging of the velocity field over the visible stellar
hemisphere, the velocities needed to produce the observed lineshifts
have to be comparable to the sound speed. Such flow velocities appear
difficult to sustain under the dissipative influence of shock
formation (e.g., Landau \& Lifschitz \cite{landau82}).

\subsection{Shock-wave propagation}

Despite the fact that full 3-dimensional hydrodynamic models with ensuing spectral-line synthesis have not yet been developed for A-type and hotter stars, various 2-dimensional atmospheric models are becoming available.  One sequence of such models, at temperatures near the onset of convection, are being developed by Holweger (\cite{holweger03}).  Models for main-sequence A-type stars of 8200 and 9000 K effective temperatures show that the atmosphere is highly dynamic. It moves up and down in a pulsation-like manner and is traversed by numerous shocks which pass up through the photosphere at supersonic velocities of typically 20--30~km~s$^{-1}$.  Rising shocks (moving towards the observer) appear blueshifted; they are accompanied by density enhancements and plausibly must contribute blueshifted spectral-line components.  Given their great amplitudes, it does not seem unreasonable that a statistical bias of the blueshifted profiles could well amount to several km~s$^{-1}$, the amounts we have been discussing.

In the absence of a detailed treatment of line formation in these models, various weighting functions can be applied on the vertical velocity field to get an indication of likely wavelength shifts for the spectral lines.  By weighting the local vertical velocity field with the local internal energy $E_i$ (a rough proxy for spectral-line excitation), a displacement is found, mimicking spectral blueshifts of 1--2~km~s$^{-1}$ (Holweger \cite{holweger03}).

A further factor is that the corresponding radial-velocity signatures probably become better visible in rapidly rotating stars.  The wavelength shift of any stellar spectrum is normally determined through a cross-correlation of the observed spectrum with some spectral template, and then the main weight comes from the flanks of the sharper stellar spectral lines.  For slowly rotating stars, many photospheric lines (originating in the lower atmosphere) contribute to the radial-velocity signal.  In rapid rotators, however, the spectral broadening causes most weak lines to become very shallow and not contributing much to the radial-velocity signal which is then weighted towards the wavelength positions of the flanks of the stronger lines (such as the Balmer lines from hydrogen).  Those typically originate in the upper photosphere, at those greater atmospheric heights where the rising shockwaves have become more fully developed, and where one could expect their blueshifted contributions to be more significant.

Even if, obviously, any genuine comparison with theory must eventually include detailed line-profile synthesis, we find this scenario the most plausible among the ones considered for explaining the intrinsic blueshifts in rapidly rotating early-type stars.  Further observational tests should include analyses of differential wavelength shifts within the spectra of individual stars, both between their weaker and stronger lines, and between the flanks and wings of stronger ones.

\section{Systematic shifts due to template mismatch}

Systematic wavelength shifts could also result from instrumental effects in the process of measuring and deducing radial velocities.  Radial-velocity instruments often use a cross correlation template taken from a slowly rotating cooler star. However, for a rapidly rotating hotter star, such a template may have a zero-point mismatch in wavelength since the rotational broadening blends different spectral-line groups than those in cooler stars (e.g., Morse et al. \cite{morse91}; Verschueren et al. \cite{verschueren99}; Griffin et al. \cite{griffin00}).

Even if effects of this type are present to some degree, we find it unlikely that they could amount to the several km~s$^{-1}$ suggested by observations (identified effects rather appear to lie on levels of 0.5--1~km~s$^{-1}$).

\section{Effects on open-cluster studies}

Irrespective of their cause -- physical or instrumental -- the existence of systematic blueshifts in some fraction of open-cluster stars can influence precise studies of cluster membership and dynamics.  For stars whose membership in a cluster is doubted, a probability of membership is normally computed, taking into account the observed deviation of the star's motion from that of the cluster as a whole.  Obviously, if for some groups of stars such measures are systematically shifted due to other effects, these stars may erroneously be rejected as cluster members (or non-members included).

Also biases may enter for any kind of studies of cluster dynamics that are based upon spectroscopic radial velocities.  The effects are likely to be greatest for groupings with many rapidly rotating stars, in particular young clusters and OB-associations.

\section{Conclusions}

Absolute and relative wavelength shifts permit certain studies of stellar atmospheric dynamics.  For some nearby, well-studied open clusters, indications are that rapidly rotating stars (especially those of early spectral types) have their spectra blueshifted by 3--5~km~s$^{-1}$.  Having searched for mechanisms that might produce such shifts, we conclude that one plausible candidate appears to be upward propagating shock waves, now seen in hydrodynamic simulations of hotter stars.  However, other mechanisms cannot be excluded, and will probably be required to explain the apparently analogous trends seen also for rapidly rotating M-type dwarfs.  Irrespective of their causes, such shifts may influence assigned cluster membership for any star, and affect the deduced cluster dynamics.  The phenomenon may also imply that rapidly rotating field stars could have underestimated radial velocities.

Future analyses of these phenomena should include, on the observational side, large and homogeneous data sets, and also separate wavelength measurements for different classes of spectral lines.  On the theoretical side, the primary need is for synthetic spectral line profiles computed from hydrodynamic model atmospheres.

\begin{acknowledgements}
This work, supported by the Swedish Research Council, The Swedish National Space Board, and the Royal Physiographic Society in Lund, has made extensive use of the open--cluster database WEBDA and the SIMBAD database at CDS.
\end{acknowledgements}

\end{document}